\documentstyle[aps,epsfig,multicol]{revtex}

\newcommand{\onecolm}{
  \end{multicols}
  \noindent\rule{0.5\textwidth}{0.1ex}\rule{0.1ex}{2ex}\hfill
}
\newcommand{\twocolm}{
  \hfill\raisebox{-1.9ex}{\rule{0.1ex}{2ex}}\rule{0.5\textwidth}{0.1ex}
  \begin{multicols}{2}
}

\begin{document}
\preprint{SNUTP 97-83}
\title{Dissipative Dynamics of Quantum Vortices\\
  in Superconducting Arrays}
\author{Mahn-Soo Choi and Sung-Ik Lee}
\address{Department of Physics, Pohang University of Science and
  Technology, Pohang 790-784, Korea}
\author{M.Y. Choi}
\address{Department of Physics, Seoul National University,
         Seoul 151-742, Korea}
\date{Aug 25 1997}
\maketitle
\draft

\begin{abstract}
We consider a two-dimensional array of ultra-small superconducting
grains, weakly coupled by Josephson junctions with large charging
energy.  We start from an effective action based on a microscopic
tunneling Hamiltonian, which includes quasiparticle degrees of
freedom, and study the resulting dissipative dynamics of quantum
vortices.  The equation of motion for a single vortex is deduced, and
compared with a commonly adopted phenomenological model.
\end{abstract}

\begin{multicols}{2}


Vortex dynamics plays an essential role in understanding the transport
properties of superconducting systems in external magnetic fields.
For instance, it is closely related to the Hall
resistivity~\cite{Hagenx93} and low-temperature magnetic
relaxation~\cite{Yeshur96}, which has drawn much attention in the
properties of vortices in two-dimensional (2D) and highly anisotropic
three-dimensional superconductors~\cite{Blatte94}.   In addition, the
dynamics of vortices has been studied extensively in superconducting
arrays as well \cite{Cerdei96}, for which the recent advancement in
fabrication techniques allows one to control the physical quantities
determining vortex dynamics, such as the vortex potential, the
effective vortex mass, and viscosity.  The superconducting arrays,
therefore, provide a convenient model system, on which various
theoretical predictions can be compared with experimental results.
Further, they may also shed light on physics of the high-temperature
ceramic superconductors, particularly in the polycrystalline form,
which behave in many respects like random arrangements of weak links.  

When the dimensions of the superconducting grains and the capacitances
involved are small, the associated charging energy is non-negligible,
and quantum dynamics of the phase comes into play at the macroscopic
level~\cite{Schonx90}.   In such an array of ultra-small junctions,
the vortices, which are defined on plaquettes of the lattice, should
be taken for quantum mechanical objects.  Here it is generally
accepted that a vortex on a superconducting array is a rather
well-defined point-like object with a finite effective mass
\cite{Eckern89,Endnote1}, and feels frictional force as well as non-dissipative
transverse force in its motion, although there has been long-standing
controversy as to the actual determination of the
latter~\cite{Hagenx93,Thoule96}.  On the other hand, a vortex
is a macroscopic object by nature, which raises the question regarding
how to macroscopically quantize the proposed classical equation of
motion for vortices in the presence of frictional force.  It is a
commonly adopted recipes for quantum dynamics of vortices to assume a
frictional force linearly proportional to the vortex velocity and then
phenomenologically quantize the resulting equation of motion according
to the Caldeira-Leggett procedure~\cite{Caldei8x}.  However, it is not
obvious that the friction should depend linearly on the vortex
velocity, and the possibility of nonlinear behavior may not be
excluded in advance.  Indeed recent numerical simulations of the
dynamics of an array containing one single vortex appears to suggest
the friction to be a nonlinear function of the vortex
velocity~\cite{Hagena94}.  This makes it necessary to investigate the
quantum dynamics of vortices based on a model closer to the first
principle, and desirable to obtain the effective action for vortices
without phenomenological presumption.

This paper presents an attempt toward such a goal: We start from an
effective action based on a microscopic model for a 2D array of
Josephson junctions consisting of ultra-small grains.   In particular,
we consider the case that the charging energy is non-negligible but
still smaller than the Josephson coupling energy, so that vortices are
well-defined, and show how the tunneling of quasiparticles introduces
dissipation to the system.  Using the dual form of the effective
action, which describes the system of dissipative quantum vortices, we
obtain the semiclassical equation of motion for a single vortex.  We
show that the damping on the vortex is  is in general nonlinear in the
vortex velocity and nonlocal in time.  However, it turns out that the
nonlinear contribution in most cases becomes negligibly small except
at very short length scales, and the frictional force in practice can
be considered to be linear in the vortex velocity, thus recovering the
commonly adopted phenomenological model.

An array of Josephson junctions can be described by
the microscopic tunneling Hamiltonian~\cite{Cohenx62}
\begin{equation}
H
= \sum_i H_i + \sum_{\langle i,j\rangle} H_{T,ij} 
  + \sum_{\langle i,j\rangle} H_{C,ij},
\end{equation}
where $H_i$ represents the microscopic Hamiltonian, e.g., the BCS
reduced Hamiltonian, for the $i$th grain.   The coupling between
neighboring islands results from the transfer of electrons through the
insulating barrier, described by $H_{T,ij}$, and from the Coulomb
interaction $H_{C,ij}$.  $H_{T,ij}$ is characterized by the Josephson
coupling energy $E_J$ whereas $H_{C,ij}$ is characterized by the
junction capacitance $C$, self-capacitance $C_0$ and the charging
energy $E_C\equiv{}e^2/2C$.   Integrating out the quasiparticle
degrees of freedom leads to a macroscopic model, where the $i$th grain
is characterized by the number $n_i$ of the superconducting electrons
(Cooper pairs) and the phase $\phi_i$ of its superconducting order
parameter~\cite{Ambega82}.  The resulting Ambegaokar-Eckern-Sch\"on
(AES) model gives the partition function in the form of a functional
integral
\begin{equation}
Z
=  \sum_{\{n_i(\tau)\}} 
   \int_0^{2\pi}{\cal D}[\phi]\;
  \exp\left[-(S_0 + S_D)\right]
  \label{eq:AES}
\end{equation}
where the Euclidean action is given by
\onecolm
\begin{eqnarray}
S_0
& = & \int_0^{\beta}d\tau\;\left[
    -i\sum_in_i\dot\phi_i
    + \frac{1}{2K}\sum_{i,j}n_i\widetilde{C}^{-1}_{ij}n_j
    - K\sum_{\langle i,j\rangle}\cos(\phi_{ij} - A_{ij})
  \right]
  \label{eq:AESa}\\
S_D
& = & \int_0^{\beta}d\tau\int_0^{\beta}d\tau'\;
  \sum_{\langle i,j\rangle}
  \alpha(\tau-\tau')\left\{
    1 - \cos\left[
      \frac{\phi_{ij}(\tau)-\phi_{ij}(\tau')}{2}
    \right]
  \right\}
  \label{eq:AESb}
\end{eqnarray}
\twocolm\noindent
with $\phi_{ij}\equiv \phi_i - \phi_j$.
Here we have rescaled the (imaginary) time in units of $1/\omega_p$,
where $\omega_p\equiv\sqrt{8E_CE_J/\hbar^2}$ is the junction plasma
frequency, and temperature in units of $\hbar\omega_p$.  Further, we
have introduced $K\equiv\sqrt{E_J/8E_C}$ and the dimensionless
capacitance matrix
\[
\widetilde{C}_{ij}
= (C_0/C + 4)\delta_{i,j}
  -\delta_{i,j+\hat{\bf x}}
  -\delta_{i,j-\hat{\bf x}}
  -\delta_{i,j+\hat{\bf y}}
  -\delta_{i,j-\hat{\bf y}}
  .
\]
The bond angle $A_{ij}$ is given by the line integral of the vector
potential due to the external magnetic field:
$A_{ij}=(2\pi/\Phi_0)\int_i^j {\bf A}\cdot d{\bf l}$, so that the
plaquette sum gives the flux per plaquette in units of the flux
quantum ($\Phi_0\equiv 2\pi \hbar c/2e$) or gauge-invariant (magnetic)
frustration, $\sum_p A_{ij}=2\pi f_{\tilde{i}}$, where $\tilde{i}$
denotes the position of the plaquette.  The quasiparticle degrees of
freedom are effectively included through the damping kernel
$\alpha(\tau)$ in the dissipative part $S_D$ given by
Eq.~(\ref{eq:AESb}).  In case that the grains are ideal BCS
superconductors with energy gap $\Delta$ and normal state resistance
$R_N$, the damping kernel is given by
\[
\alpha(\tau)
= \frac{\Delta^2}{R_N}K_1^2(\Delta|\tau|)
\]
in the low-temperature limit ($T\to 0$), where $K_1$ is the modified
Bessel function and we rescaled the gap energy $\Delta$ and normal
state resistance $R_N$ by $\Delta/\hbar\omega_p\to\Delta$ and
$R_N/R_0\to{}R_N$, respectively.  It should be stressed here that the
damping term entirely originates from the intergrain quasiparticle
tunneling and includes neither the effects of the Cooper pair decay
into the pool of normal electrons nor those of Ohmic shunt between
grains.

A variety of properties of Josephson-junction systems have been
successfully described by the effective action in Eqs.\
(\ref{eq:AESa}) and (\ref{eq:AESb}).  We thus employ the AES model as
a good starting point for an effective model for the dissipative
dynamics of quantum vortices.  For this purpose, it is convenient to
use the dual transformation, which rewrites the model in terms of the
vortex variables instead of the original charge (Cooper pair)
variables~\cite{Josexx77,vanWee91,MYChoi94}.   In the presence of the
quasiparticle dissipation, the resulting effective model
reads\cite{Faziox92}
\begin{equation}
Z
= \sum_{\{n^v\}}\int{\cal D}[\phi^v]\;
  \exp\left[ -(S_0^v+S_D^v) \right]
\label{eq:AES.vortex:a}
\end{equation}
with
\onecolm
\begin{equation}
S_0^v
= \int_0^{\beta}d\tau\;\left[
    - i\sum_in^v_i\dot\phi^v_i
    + 2\pi^2K\sum_{i,j}(n^v_i - f_i)G_{ij}(n^v_j-f_j)
    - \sum_{\langle i,j\rangle}
      \frac{1}{4\pi^2K}\cos\phi^v_{ij}
  \right]
  \label{eq:AES.vortex:b}
\end{equation}
\begin{equation}
S_D^v
= \int_0^{\beta}d\tau\int_0^{\beta}d\tau'\;
  \sum_{\langle i,j\rangle}\alpha(\tau-\tau')\left\{
    1-\cos\left[
      \frac{\theta_{ij}(\tau)-\theta_{ij}(\tau')} {2}
    \right]
  \right\}
  ,
\label{eq:AES.vortex:c}
\end{equation}
\twocolm\noindent
where $i, j$ now denote the dual lattice sites, i.e., the tilde signs
representing the dual lattice sites
have been dropped for simplicity. 
The vortex charge (in units of $\Phi_0$) 
$n^v_i$ and the vortex phase $\phi^v_i$
(of the macroscopic vortex wave function) are conjugate to each other,
$G_{ij}$ is the lattice Green's function, and
$\theta_i\equiv -2\pi\sum_j G_{ij}n_j^v$.

We now investigate the dynamics of a single vortex, focusing on the
nature of the dissipation on the vortex.  Such a single vortex may
actually be introduced by adjusting weak external magnetic field at
zero temperature.  For a vortex at position ${\bf r(\tau)}$, the
dissipative action takes the form
\begin{equation} 
S_D^{v}
= \int_0^\beta{}d\tau\int_0^\beta{}d\tau'\;
  \alpha(\tau-\tau')
  W\left({\bf r}(\tau)-{\bf r}(\tau')\right),
\label{eq:SDvSingle}
\end{equation}
where the non-local self interaction
\onecolm
\begin{equation} 
W\left({\bf r}(\tau)-{\bf r}(\tau')\right)
\equiv \sum_{<{\bf r}{\bf r}'>} \left[
     1 - \cos\left(
       \frac{
         \theta_{{\bf r}{\bf r}'}(\tau)
         -\theta_{{\bf r}{\bf r}'}(\tau')
       }{2}
     \right)
  \right]
\end{equation}
can be evaluated to reveal the logarithmic dependence
\begin{equation}
W\left({\bf r}(\tau)-{\bf r}(\tau')\right)
\approx \frac{\pi}{2}\ln\left[
    \frac{{\bf r}(\tau)-{\bf r}(\tau')}{a}
  \right]
  ,
\end{equation}
\twocolm\noindent
aside from an irrelevant additive constant.
The procedures of action minimization and analytic 
continuation~\cite{Ambega82} then lead to the semiclassical
equation of motion describing the real-time dynamics of a single
vortex driven by an applied current $I$ (in units of the Josephson
critical current $I_J$) in the $x$-direction
\onecolm
\begin{equation}
2\pi^2K\ddot{\bf{r}} 
- 2\int^tdt'\;
  \alpha(t-t')\nabla_{\bf{r}}W({\bf{r}}(t)-{\bf{r}}(t'))
= 2\pi{}KI\hat{\bf{u}}\times\hat{\bf{z}}
  .
\label{eq:eqofmotion:a}
\end{equation}
\twocolm\noindent
The damping kernel $\alpha(t)$ in the real-time domain 
may be obtained by the analytic continuation
\begin{equation}
\alpha(t)
= \int_{-\infty}^\infty\frac{d\omega}{2\pi}\;
  e^{-i\omega k}\alpha(i\omega_n\to\omega +i 0^+)
  .
\end{equation}
Alternatively, $\alpha(t)$ can also be obtained directly through the use of
the real-time formalism~\cite{Caldei8x,Schmid82}, which, for the actions in
Eqs.~(\ref{eq:AESa},\ref{eq:AESb}), gives $\alpha(t)$ related
to the quasiparticle tunneling current $I_{\rm qp}(\omega)$
(in units of $I_J$) via \cite{Ambega82,Wertha66}
\begin{equation}
i\alpha(\omega)
= KI_{\rm qp}(\omega)
  . 
  \label{eq:alpha-Iqp}
\end{equation}

Equation (\ref{eq:eqofmotion:a}) is the most general semiclassical
equation of motion at length scales larger than the lattice constant.
At zero temperature ($T=0$), the quasiparticle tunneling current is
simply given by
\begin{equation}
\tilde{I}_{\rm qp}(\omega)
= \left\{ \begin{array}{ll}
    0, & |\tilde{\omega}|<2\Delta\\
    2\gamma\omega, & |\tilde{\omega}|>2\Delta
  \end{array} \right. \label{quasi}
  ,
  \label{IqpT0}
\end{equation}
where $\gamma\equiv{}1/\omega_pR_NC$.  This allows one to
write the equation of motion in a more explicit and appealing form.
The damping kernel in this case reduces to
\begin{equation}
\label{eq:alpha}
\alpha(t)
= 2\gamma{}K\frac{d}{dt}\left[
    \delta(t)-\Delta{\rm{sinc}}(2\Delta{}t)
  \right]
  ,
\end{equation}
where ${\rm{}sinc}x\equiv(2/\pi{}x)\sin{x}$.
Here it should be noted that the short time behavior of Eq.\
(\ref{eq:alpha}) is valid only approximately, because of the high
frequency cutoff in $\alpha(\omega)$.  
With this simple form of the damping kernel, we finally obtain 
\onecolm
\begin{equation}
\ddot{\bf{r}}
+ \gamma\dot{\bf{r}}
- \gamma\frac{\Delta}{\pi^2}
  \int^tdt'\;{\rm{sinc}}[2\Delta(t-t')]
  W''({\bf{r}}(t)-{\bf{r}}(t'))\dot{\bf{r}}(t')
= - \frac{1}{\pi}I\hat{\bf{y}}
  ,
\label{eq:eqofmotion:b}
\end{equation}
which, at long time scales, Eq.~(\ref{eq:eqofmotion:b}) takes the more
explicit form 
\begin{equation}
\ddot{\bf{r}}
+ \gamma\dot{\bf{r}}
- \gamma\Delta
  \int^tdt'\;{\rm{sinc}}[2\Delta(t-t')]
  \frac{\dot{\bf{r}}(t')}{\pi|{\bf{r}}(t)-{\bf{r}}(t')|^2}
= - \frac{1}{\pi}I\hat{\bf{y}}
  .
\label{eq:eqofmotion:c}
\end{equation}
\twocolm\noindent

\begin{figure}
\centerline{\epsfig{file=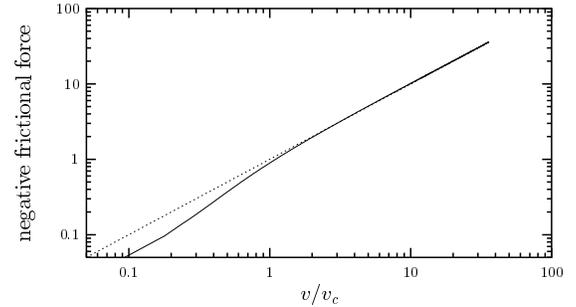,clip=,width=85mm}}
\caption{Behavior of the frictional force (in arbitrary units) as a
  function of the vortex velocity.  The dotted line represents the
  usual linear frictional force.  The logarithmic scale should be
  noticed.  The values of the parameters are $R_N=10.0$ and
  $K\Delta=1/8$.}
\label{fig:f(v)}
\end{figure}

The semiclassical equation of motion given by Eq.\
(\ref{eq:eqofmotion:b})
or Eq.\ (\ref{eq:eqofmotion:c}) possesses two damping terms: One is
linear, but the other is 
nonlinear in the vortex velocity and nonlocal (memory-dependent) in time.
When the vortex velocity is large, we have
$\dot{\bf r}(t')/|{\bf r}(t)-{\bf r}(t')|^2\sim 1/v(t')$ in Eq.\
(\ref{eq:eqofmotion:c}),
and the nonlinear term becomes sufficiently small compared with the
ordinary linear friction term.  For small velocities, on the other
hand, one must be careful about the short-wavelength
cutoff~\cite{Eckern90} present in the function $W({\bf r})$.   (Note
that the continuum approximation has been used in the derivation of
the equation of motion.) To examine the behavior at low velocities, we
have thus numerically integrated the equation of motion, and display
the obtained frictional force as a function of the vortex velocity in
Fig.\ \ref{fig:f(v)}.  It is revealed that the frictional force
slightly deviates from the linear behavior at velocities smaller than
$v_c\equiv\Delta$.   It is of interest to note that $v_c$
may be written in the form $v_c=a\omega_J/\pi$ (in natural unit),
which corresponds to vortex hopping by one lattice constant during the
characteristic time $1/\omega_J\equiv\hbar/2|e|R_NI_J$ associated with
the Josephson oscillation in a resistively shunted Josephson junction.

The above analysis demonstrates that the frictional force on a
vortex is {\em practically} linear in the vortex velocity, in particular
in the long-time and long-wavelength scale, 
where the semiclassical equation of motion is mostly concerned.
Neglecting the nonlinear friction and 
rescaling the time in units of $1/\omega_J$,
we have Eq.\ (\ref{eq:eqofmotion:c}) in the reduced form
\begin{equation}
\pi\beta_c\ddot{\bf r}
+ \pi\dot{\bf r}
= -I\hat{\bf y}
  ,
\end{equation}
which precisely corresponds to the commonly adopted phenomenological
equation of motion describing the 
resistively and capacitively shunted junction
(RCSJ) model with the Stewart-McCumber parameter
$\beta_c\equiv\omega_JR_NC$\cite{Eckern89}.
This is remarkable in view of the fact that we have considered only 
quasiparticle tunneling, and neither the Ohmic shunt nor
any other local damping sources have been included.

In conclusion, we have considered a microscopic model for a
two-dimensional array of Josephson junctions, including the
quasiparticle degrees of freedom.  From the effective
action, which has been obtained without any phenomenological
presumptions, the semiclassical equation of motion for a single vortex
has been deduced.   It has been revealed that the quasiparticle
tunneling produces friction on the vortex motion.  It includes a
nonlinear term in addition to the ordinary linear term although the
nonlinear friction is in most cases dominated by the latter.  At
finite temperatures, the quasiparticle tunneling current $I_{\rm
qp}(\omega)$ is smoothed out and approaches the Ohmic behavior (i.e.
$I_{\rm qp}(\omega)\propto\omega$).  In consequence, the non-linear
friction term would become even less pronounced and essentially
negligible.  Similarly, arrays of Josephson junctions between $d$-wave
superconductors are expected to display essentially linear behavior
since $d$-wave superconductors have nodes in the momentum space at
which the energy gap vanishes.   This leads to the quasiparticle
tunneling current with no sharp threshold in $\omega$, even at $T=0$,
separating the high frequency Ohmic behavior and the low frequency
non-Ohmic behavior~\cite{Mandru93}.  It is of interest to compare the
nonlinear behavior found in this work with that obtained in
Ref.~\cite{Hagena94}.  In the latter, numerical integration of the
phenomenological RCSJ model has led to effective damping in the vortex
motion, which becomes nonlinear as the velocity increases.  Here,
unlike the existing phenomenological approach, we have started from a
microscopic model, and derived explicitly the equation of motion for a
vortex in the system.  The resulting damping term displays
nonlinearity mainly in the low-velocity regime, which is in contrast
with that obtained in the phenomenological approach.
In addition to the quasiparticle tunneling, spin-wave excitations may
provide another mechanism for the damping of vortex
motion~\cite{Faziox94,Eckern93}.  Since
the spin-wave excitation requires energy of order of
$\hbar\omega_p$~\cite{Eckern93} for $\hbar\omega_p\gtrsim\Delta$, it is rather irrelavant.
Furthermore, in the case of discrete charge states considered here, the region where the spin-wave damping can be
ignored gets wider~\cite{Faziox94}.  Nevertheless for
$\hbar\omega_p\ll\Delta$, however, contributions of spin-wave
excitations may not be disregarded, and the investigation of the
dissipation due to both the quasiparticle and spin-wave
excitations will be a challenging topic.  Finally, we
remark that we have considered only the single vortex motion.  In the
case of many vortices, the vortex-vortex interaction becomes crucial,
especially for $C_0\ll C$, which yields the logarithmically long
interaction range.  The interaction effects, based on the highly
suggestive effective action in Eqs.\ (\ref{eq:AESa}) and
(\ref{eq:AESb}), will be an interesting topic for further study.

This work was supported in part by the Basic Science Research
Institute Program, Ministry of Education of Korea and by the Korea
Science and Engineering Foundation through the SRC Program (MYC).
MSC also acknowledges the financial aid from the Seoam Scholarship
Foundation.


\begin{thebibliography}{10}

\bibitem{Hagenx93}
S.~J. Hagen, Phys. Rev. B {\bf 47},  1064  (1993), and references therein.

\bibitem{Yeshur96}
Y. Yeshurun, A.~P. Malozemoff, and A. Shaulov, Rev. Mod. Phys. {\bf 68},  911
  (1996).

\bibitem{Blatte94}
G. Blatter {\it et~al.}, Rev. Mod. Phys. {\bf 66},  1125  (1994).

\bibitem{Cerdei96}
 See, e.g., Physica B {\bf 222 (4)}, 253--406 (1996).

\bibitem{Schonx90}
G. Sch\"{o}n and A.~D. Zaikin, Phys. Rep. {\bf 198},  237  (1990).

\bibitem{Eckern89}
U. Eckern and A. Schmid, Phys. Rev. B {\bf 39},  6441  (1989), and references
  therein.

\bibitem{Endnote1}
This still remains controversial for continuum systems. See, for example, J.-M.
  Duan and A.~J. Leggett, Phys. Rev. Lett. {\bf 68}, 1216 (1992); Q. Niu, P.
  Ao, and D.~J. Thouless, Phys. Rev. Lett. {\bf 72}, 1706 (1994); {\bf 75}, 975
  (1995); J.-M. Duan, Phys. Rev. Lett. {\bf 75}, 974 (1995).

\bibitem{Thoule96}
D.~J. Thouless, P. Ao, and Q. Niu, Phys. Rev. Lett. {\bf 76},  3758  (1996);
  G.~E. Volovik, Phys. Rev. Lett. {\bf 77}, 4687 (1996).

\bibitem{Caldei8x}
A.~O. Caldeira and A.~J. Leggett, Phys. Rev. Lett. {\bf 46},  211  (1981);
  A.~O. Caldeira and A.~J. Leggett, Ann. Phys. {\bf 149}, 374 (1983).

\bibitem{Hagena94}
T.~J. Hagenaars, P.~H.~E. Tiesinga, J.~E. van Himbergen, and J.~V. Jos\'e,
  Phys. Rev. B {\bf 50},  1143  (1994).

\bibitem{Cohenx62}
M.~H. Cohen, L.~M. Falicov, and J.~C. Phillips, Phys. Rev. Lett. {\bf 8},  316
  (1962).

\bibitem{Ambega82}
V. Ambegaokar, U. Eckern, and G. Sch\"{o}n, Phys. Rev. Lett. {\bf 48},  1745
  (1982); U. Eckern, G. Sch\"{o}n, and V. Ambegaokar, Phys. Rev. B {\bf 30},
  6419 (1984).

\bibitem{Josexx77}
J.~V. Jos\'e, L.~P. Kadanoff, S. Kirkpatrick, and D.~R. Nelson, Phys. Rev. B
  {\bf 16},  1217  (1977); R. Savit, Rev. Mod. Phys. {\bf 52}, 453 (1980).

\bibitem{vanWee91}
B.~J. van Wees, Phys. Rev. B {\bf 44},  2264  (1991).

\bibitem{MYChoi94}
M.~Y. Choi, Phys. Rev. B {\bf 50},  10088  (1994); 13875 (1994).

\bibitem{Faziox92}
R. Fazio {\it et~al.}, Helv. Phys. Acta {\bf 65},  228  (1992).

\bibitem{Schmid82}
A. Schmid, J. Low Temp. Phys. {\bf 49},  609  (1982).

\bibitem{Wertha66}
N.~R. Werthamer, Phys. Rev. {\bf 147},  255  (1966).

\bibitem{Eckern90}
U. Eckern,  in {\em Applications of Statistical and Field Theory Methods to
  Condensed Matter}, edited by D. Baeriswyl {\it et~al.} (Plenum Press, New
  York, 1990).

\bibitem{Mandru93}
D. Mandrus {\it et~al.}, Europhys. Lett. {\bf 22},  199  (1993).

\bibitem{Faziox94}
R. Fazio, A. van~der Otterlo, and G. Sch\"on, Europhys. Lett. {\bf 25},  453
  (1994).

\bibitem{Eckern93}
U. Eckern and E.~B. Sonin, Phys. Rev. B {\bf 47},  505  (1993); U.
  Geigenm\"uller, C.~J. Lobb, and C.~B. Whan, Phys. Rev. B {\bf 47}, 348
  (1993); P.~A. Bobbert, Phys. Rev. B {\bf 45}, 7540 (1992).

\end{thebibliography}

\end{multicols}

\end{document}